\documentclass[fleqn,12pt,twoside]{article}
\usepackage{espcrc1}
\usepackage{epsf}
\usepackage{graphicx}

\newcommand{\be}{\begin{eqnarray}}
\newcommand{\ee}{\end{eqnarray}}

\newcommand{\AmS}{{\protect\the\textfont2
  A\kern-.1667em\lower.5ex\hbox{M}\kern-.125emS}}

\title{Dissipation near the QCD phase transition}

\author{\'Agnes M\'ocsy\address[MCSD]{The Niels Bohr Institute, \\
        Blegdamsvej 17, 2100 Copenhagen, Denmark}%
      }

\begin{document}

\maketitle

\begin{abstract}
We set up a framework for field theoretical studies of systems out of thermal 
equilibrium and zoom in on the dissipation of disoriented chiral condensates. 
Short relaxation times are obtained in the phase transition region, 
jeopardizing the definiteness of a DCC signal.
\end{abstract}

\section{Introduction}

QCD at finite temperature $T$ and baryon density $n_B$ is really hard to be 
tackled directly, except in two limiting cases: At very high $T$ and/or $n_B$
the coupling is weak allowing for perturbative calculations. At very low 
 $T$ and/or $n_B$ the theory is strongly coupled, and chiral 
perturbation theory proved to provide a good description. In the 
nonperturbative region between these two limits effective field theories are 
needed. Such theories can be proved or disproved by experiments and lattice 
QCD simulations. Effective models incorporate only certain aspects of QCD, 
such as chiral symmetry. There is a phase transition between the above two 
limits, corresponding to symmetry restored and symmetry broken phases.
We believe that as the matter produced at RHIC-Brookhaven National 
Laboratory \cite{rhic} in the collision of ultrarelativistic heavy ions 
expands, it cools through this QCD phase transition. Experimental 
observations \cite{brams} suggest that the produced matter is almost baryon 
free. In the following, we discuss hot baryon-free matter. Cold dense quark 
matter phases are discussed in \cite{rajagopal}.

There are a large amount of models describing the chiral phase transition. 
The majority of these assume thermal equilibrium. RHIC results suggest though 
\cite{star}, that even if thermal equilibrium is achieved in some early 
stages of the collision, most probably it is not maintained as the matter 
rapidly expands and the temperature quickly drops below the transition 
temperature. The dynamical evolution of such an out-of-equilibrium system is 
not yet understood. When it comes to possible approaches, lattice simulations 
unfortunately do not prove to be the way to go, since lattice can describe 
static situations in thermal equilibrium. Therefore, one has to reside to 
different theoretical approaches. 

A transition from ordered to disordered phase is accompanied by the formation 
of condensates. Here we talk about the chiral condensate, 
$\Phi=(\sigma,\vec{\pi})$, parametrized using the variables of the sigma 
model. In the phase where chiral symmetry is restored the condensate 
vanishes $\Phi=(0,\vec{0})$. As soon as the system passes through the phase 
transition chiral symmetry is broken and the condensate evolves back into its 
true ground state $\Phi=(f_\pi,\vec{0})$. The temporary restoration of chiral 
symmetry may result in the formation of disoriented chiral condensates (DCC) 
\cite{rw}: As the system passes through the phase transition configurations 
in which the condensate is oriented differently than the normal ground state 
can develop, $\Phi=(v,\vec{\pi}\neq 0)$. DCCs relax to the correct ground 
state through the emission of low momentum pions with an anomalous 
distribution in isospin space. Detecting fluctuations in the ratio of 
produced neutral pions compared to the charged ones can serve as a signal of 
DCC production. DCC signals have not been observed at CERN-SPS. The STAR 
detector at RHIC searches for dynamical fluctuations on an event-by-event 
basis and may be therefore better suited for DCC searches. 
 
The ability to detect DCCs depends on their lifetime. The original work 
\cite{rw} assumes a total quench scenario, meaning that at the critical 
temperature long-wavelength chiral fields completely decouple from the 
thermal background. In heavy ion collisions production of pions with high 
momenta is significant \cite{brams}. The effect of these thermal degrees 
of freedom should not be neglected. If energy exchange between the condensate 
and the heat bath becomes possible decay channels open up and scattering 
occurs. These processes are responsible for the dissipation of the 
non-equilibrium condensate, and thus the reduction of the lifetime of DCCs. 
Early estimates predicted $1~$fm/c \cite{biro}. More recent calculations 
yield $4-7~$fm/c \cite{steele}. We determine the lifetime of DCCs formed 
within a heat bath of mesons \cite{mocsy}, developing a consistent 
semi-classical description of the chiral condensate out of thermal 
equilibrium. Some other related works are \cite{rischke,cejk,bettencourt}.

\section{The theory}

For studying DCCs the linear sigma model proved to be convenient framework. 
We decompose the fields into a non-equilibrium condensate and fluctuations 
about it. We do this by separating the different Fourier components of the 
fluctuation introducing a momentum scale $\Lambda_c$
\be 
\Phi(x)=\bar\Phi+\Phi_s(x)+\Phi_f(x)\, . 
\ee
$\Phi_s=(\sigma_s,\vec{\pi}_s)$ are the long wavelength modes of the chiral 
order parameter, $\langle\Phi\rangle=\bar\Phi+\Phi_s$, slowly varying 
fluctuations, representing modes with momentum $\mid\vec{k}\mid<\Lambda_c~$. 
These soft modes are occupied by a large number of particles and may then be 
treated as classical fields. $\Phi_f=(\sigma_f,\vec{\pi}_f)$ are high 
frequency modes with $\mid\vec{k}\mid>\Lambda_c~$. These hard modes, 
represent quantum and thermal fluctuations, and constitute a heat bath. By 
having nonzero pion condensate, $\vec{\pi}_s\neq 0$, we allow for the 
presence of DCCs. The equilibrium chiral condensate is chosen to lie along 
the sigma direction $\bar{\Phi} = (v,\vec{0})$.

We have derived classical effective equations for the non-thermal condensate, 
$\Phi_s$, embedded in the thermal background. The effect of $\Phi_f$ is taken 
into account perturbatively, and improved with resummation. We coarse-grained 
these equations, which means that we averaged these over time and length 
scales that are short compared to the scales characterizing the slow fields, 
but long relative to the scales of the quantum and thermal fluctuations. 
Accordingly, $\langle\Phi_f\rangle = 0$, but quadratic terms and 
cross-correlations are nonzero. The non-equilibrium fluctuations we define as 
the sum of equilibrium fluctuations and deviations from this:
\be
\langle\Phi_f^i\Phi_f^j\rangle = \langle\Phi_f^i\Phi_f^j\rangle_{eq}
\delta_{ij} + \delta(\Phi_f^i\Phi_f^j)\, ,
\ee
where $\delta(\Phi_f^i\Phi_f^j)$ are the responses of the heat bath to the 
presence of the condensate. Assuming the system is slightly out of 
equilibrium, $\delta(\Phi_f^i\Phi_f^j)$ are determined using linear response 
theory.
The presence of linear response functions in the equations of motion is 
crucial: They renormalize the meson masses, change the velocity of 
propagation of the soft modes, and also introduce dissipation. Further details 
are found in \cite{mocsy}. 

\section{Results}

\begin{figure}[t]
\begin{minipage}[t]{77mm}
\leavevmode
\parbox{8.cm}{%
\epsfxsize=8.cm
\epsffile{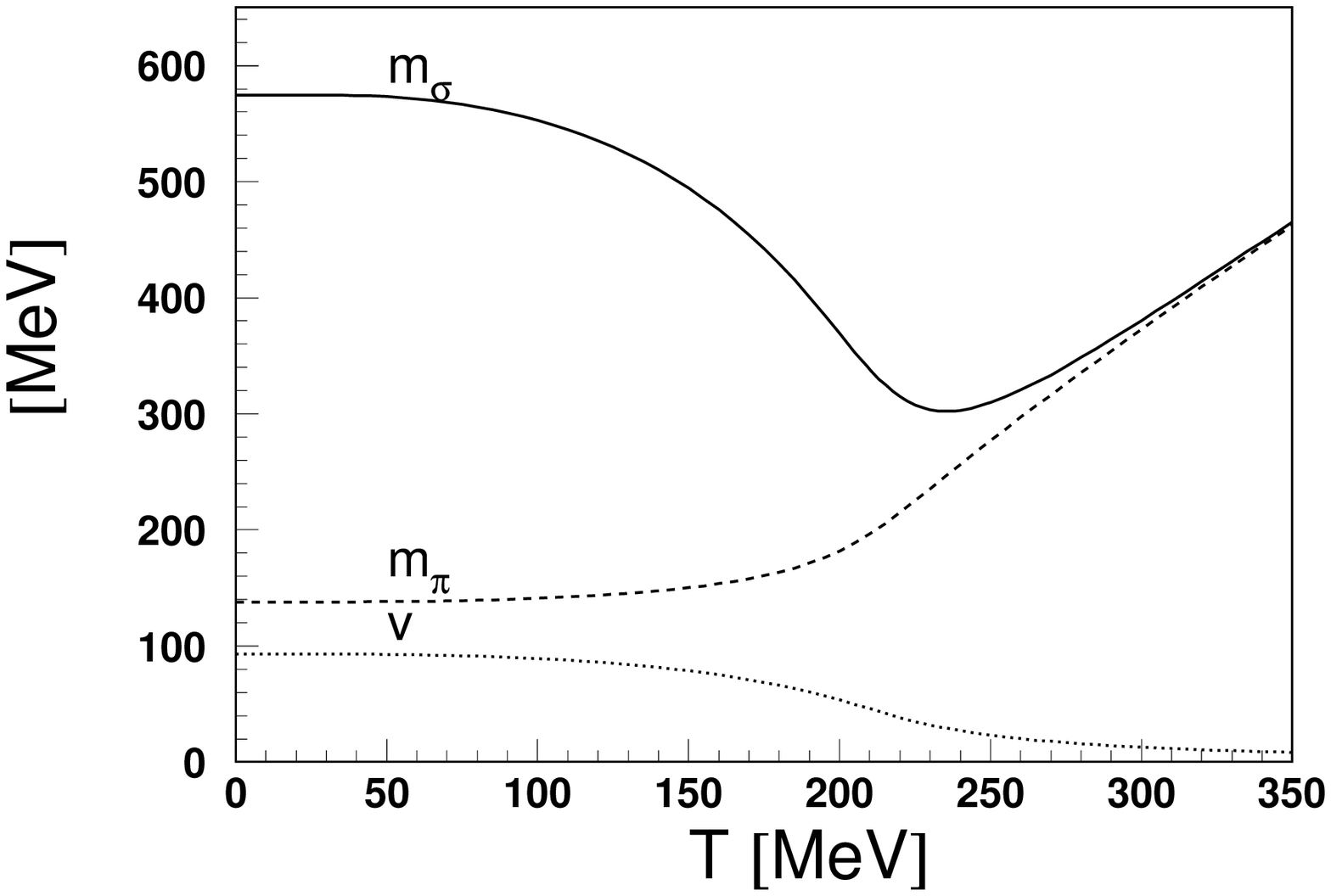}
}
\vspace*{-1cm}
\caption{Temperature dependence of the resummed meson masses and of the 
equilibrium condensate.}
\label{mass.fig}
\end{minipage}
\hspace{\fill}
\begin{minipage}[t]{77mm}
\leavevmode
\parbox{7.8cm}{%
\epsfxsize=7.6cm
\epsffile{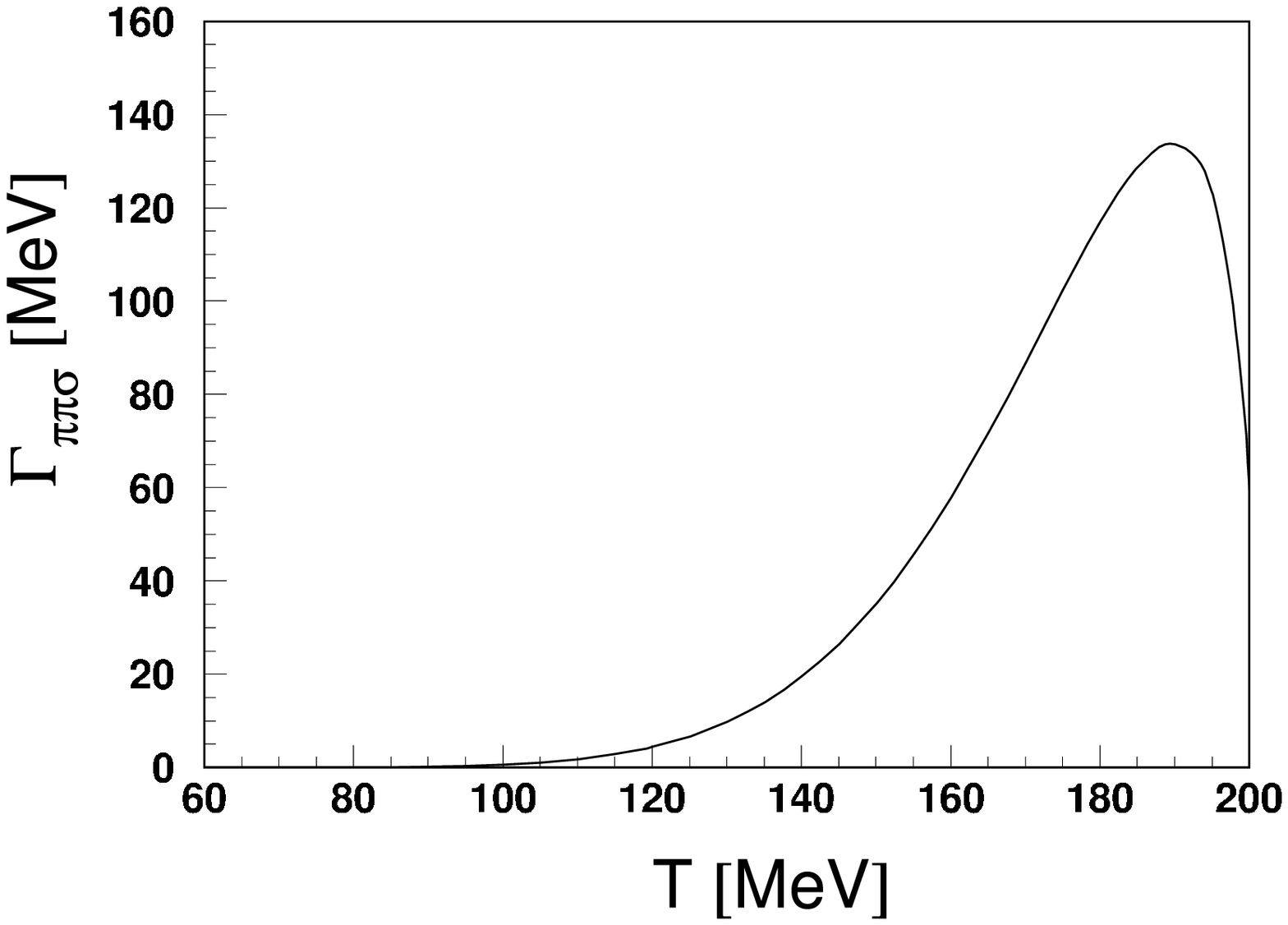}
}
\vspace*{-1cm}
\caption{Temperature dependence of the pion damping rate due to 
$\pi_s\pi_f\rightarrow\sigma_f$.}
\label{one.fig}
\end{minipage}
\end{figure}
One-loop self-consistent numerical solutions for the meson masses and the 
equilibrium condensate are shown in figure \ref{mass.fig}. The equilibrium 
condensate monotonically decreases with increasing temperature. A crossover 
region can be defined where the sigma and pion masses start to approach 
degeneracy, signalling approximate symmetry restoration.  

\begin{figure}[t]
\begin{minipage}[t]{77mm}
\leavevmode
\hbox{%
\epsfxsize=8.1cm
\epsffile{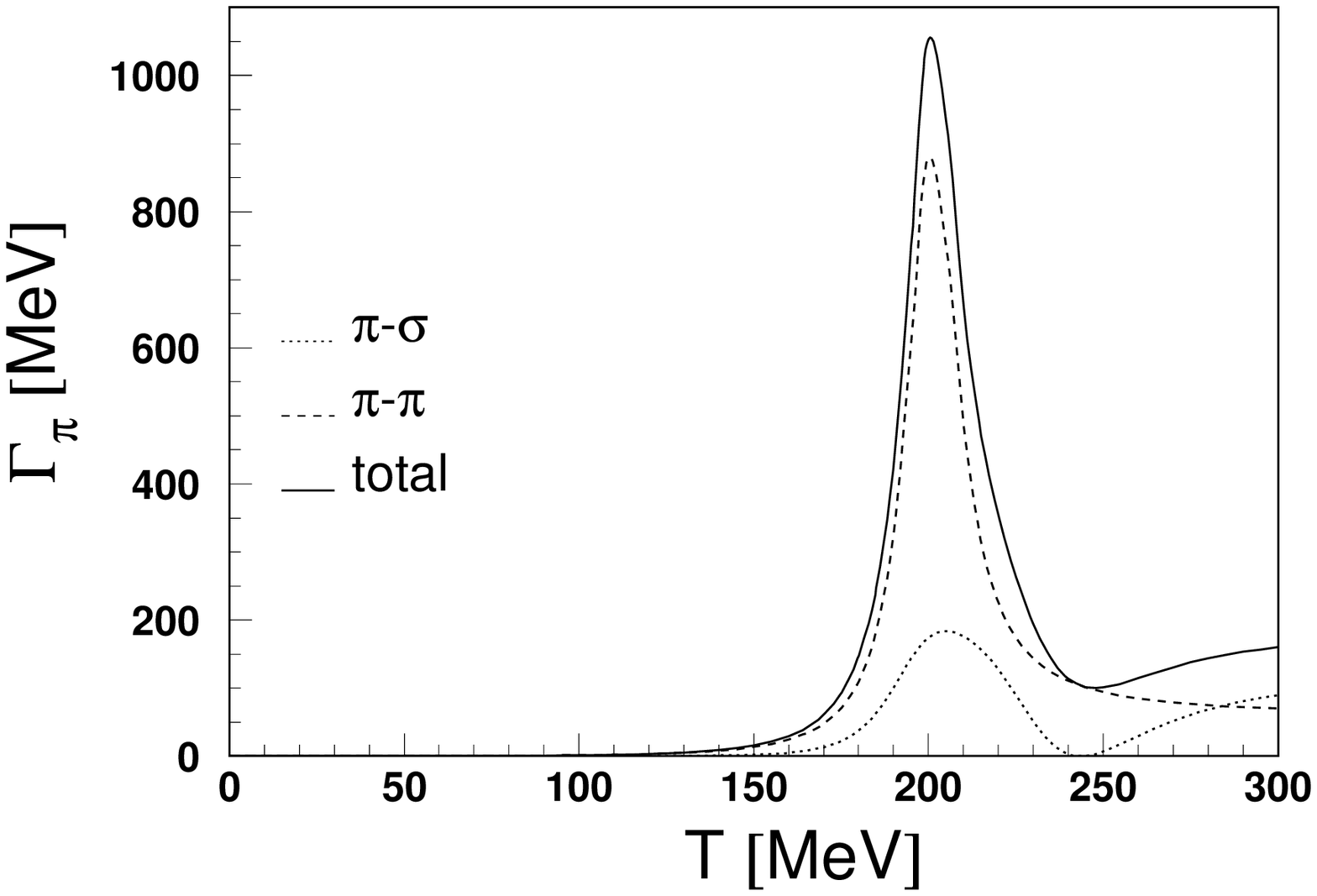}
}
\vspace*{-1cm}
\caption{Temperature dependence of the pion damping due to 
$\pi_s\pi_f\rightarrow\pi_f\pi_f$ and $\pi_s\sigma_f\rightarrow\sigma_f\pi_f$ 
scattering.}
\label{two.fig}
\end{minipage}
\hspace{\fill}
\begin{minipage}[t]{77mm}
\leavevmode
\hbox{%
\epsfxsize=7.6cm
\epsffile{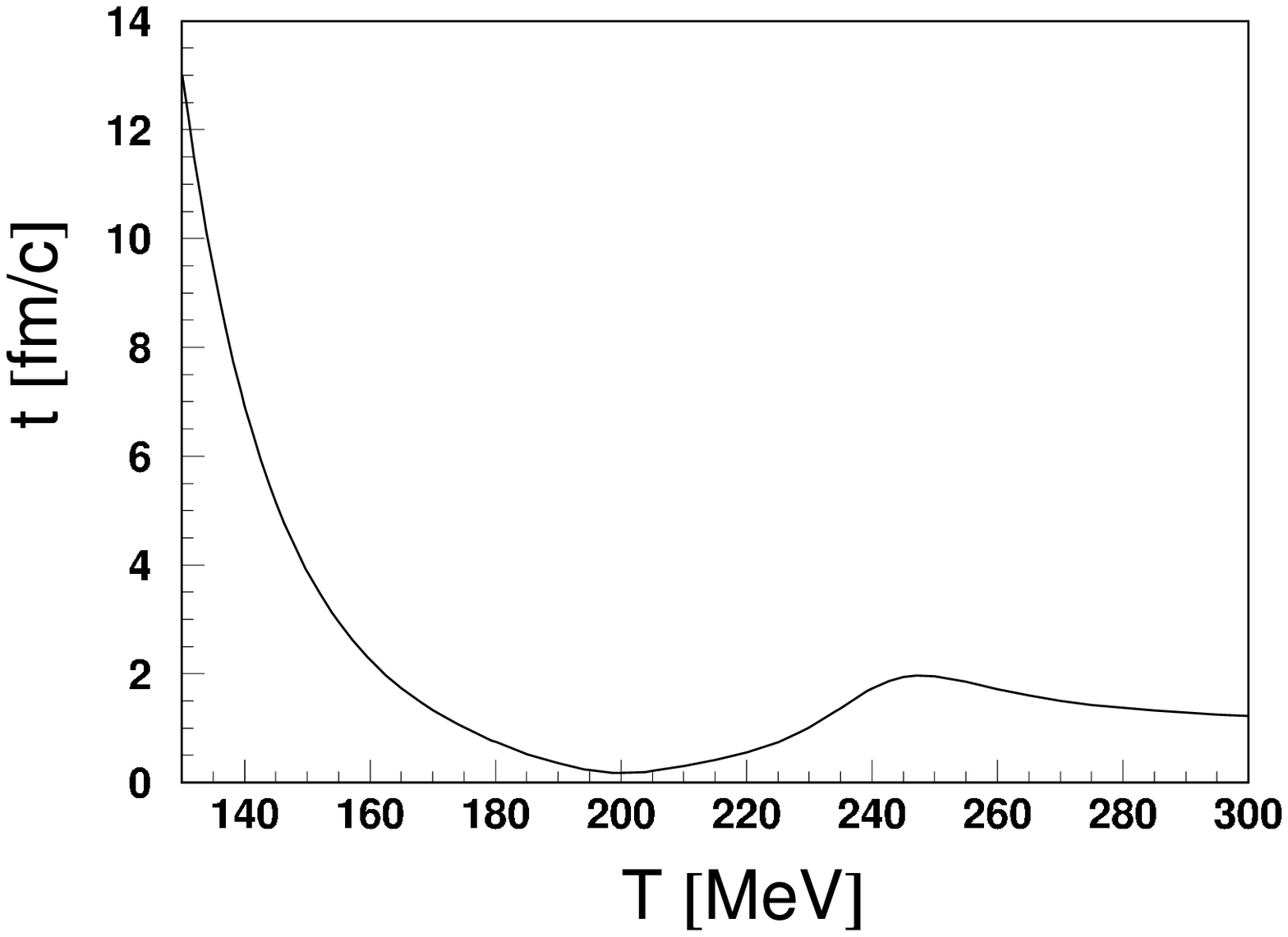}
}
\vspace*{-1cm}
\caption{Temperature dependence of the relaxation time of homogeneous 
condensates.}
\label{time.fig}
\end{minipage}
\end{figure}

A soft pion from the condensate can annihilate with a hard thermal pion 
producing a hard thermal sigma meson, $\pi_s\pi_f\rightarrow\sigma_f$, 
provided $m_\sigma\geq 2m_\pi$. The temperature dependence of the net rate of 
dissipation, $\Gamma_{\pi\pi\sigma}$, is shown in figure \ref{one.fig}. At 
$T=0$ the dissipation is zero, because this is exclusively a finite 
temperature process. At low $T$ the available phase space is suppressed by 
the large sigma mass. With increasing T the sigma mass is dropping and the 
width of the pions is increasing. At $T\simeq 170~$MeV, for example, the 
damping is about $55\%$ of its energy. Dissipation of DCCs can also arise 
from scattering of soft pions with thermal pions 
$\pi_s\pi_f\rightarrow\pi_f\pi_f$, or thermal sigmas, 
$\pi_s\sigma_f\rightarrow\sigma_f\pi_f$. The damping rates are shown in 
figure \ref{two.fig}. Again, at low T there is a strong suppression due to 
the heavy sigma exchange. With increasing T the scattering rate, 
$\Gamma_\pi$, grows comparable to $\Gamma_{\pi\pi\sigma}$. In the critical 
region the contribution from pion-pion scattering grows rapidly reaching a 
maximum at $T=200~$MeV. This is the same temperature as where 
$\pi_s\pi_f\rightarrow\sigma_f$ becomes forbidden by the kinematics.  

The rate at which equilibrium is approached is directly controlled by the 
damping through the relation
\be
t = \frac{1}{\Gamma_{\pi\pi\sigma}+\Gamma_\pi}\,~,
\ee
and is presented in figure \ref{time.fig}. In the phase transition region the 
decay time is the smallest. At $T=200~$MeV we found $t=0.17~$fm/c, and at 
$T=235~$MeV $t=1.36~$fm/c. When assuming no multiple interactions with the 
heat bath then $t$ is the lifetime of the DCC. The times we obtained are 
shorter than previous estimates \cite{biro,steele}, and are short enough to 
make possible DCC signals questionable. One can expect that multiple 
scatterings would further increase the damping, thus further decreasing the 
relaxation time.

\section{Summary and conclusions}

Motivated by the survival possibility of DCCs in the background of a multitude 
of thermal particles formed after two heavy ions are collided at
ultrarelativistic energies, we calculated the damping of the non-thermal 
chiral condensate. We found that not only decay but also scattering processes 
are significant dissipation sources. In the phase transition region short 
relaxation times were obtained, which makes the observation of a possible 
DCC signal questionable. 


\end{document}